\documentstyle[aps,preprint,epsfig]{revtex}
\tighten
 \begin{document}
\draft
\newpage
\setcounter{page}{1}
\title{Bose-Einstein condensation in dense nuclear matter and 
strong magnetic fields}
\author{Prantick Dey$^{\rm (a)}$, Abhijit Bhattacharyya$^{\rm (b)}$ and Debades 
Bandyopadhyay$^{\rm (c)}$}
\address{$^{\rm (a)}$Physics Department, Raja Peary Mohan College, Uttarpara, 
Hooghly 712258, India} 
\address{$^{\rm (b)}$Physics Department, Scottish Church College, 1 \& 3
Urquhart Square, Calcutta 700006, India} 
\address{$^{\rm (c)}$Saha Institute of Nuclear Physics, 1/AF Bidhannagar, 
Calcutta 700 064, India}
\maketitle
\vspace{1cm}

\begin{abstract}
Bose-Einstein condensation of antikaons in cold and dense beta-equilibrated 
matter under the influence of strong magnetic fields is studied within a 
relativistic mean field model. For magnetic fields $> 5 \times 10^{18}$G, 
the phase spaces of charged particles are modified resulting in
compositional changes in the system. The threshold density of $K^-$ 
condensation is shifted to higher density compared with the field free case. 
In the presence of strong fields, the equation of state becomes stiffer than 
that of the zero field case. 
\end{abstract}

\newpage
Recently, it has been inferred that some soft gamma ray repeaters (SGRs) and 
perhaps certain anomalous X-ray pulsars (AXPs) could be neutron stars 
having large magnetic fields $\sim 10^{14}-10^{16}$ G \cite{Kou98}. Those 
objects are called "magnetars" \cite{Dun92}. Earlier large magnetic fields 
$\sim 10^{13}$G were estimated to be associated with the surfaces of some radio 
pulsars \cite{Chan}. The origin of such ultra strong magnetic fields is 
still an unsolved problem. An attractive idea about the origin is that the
small magnetic field of a progenitor star is amplified due to the magnetic
flux conservation during the gravitational collapse of the star 
\cite{Chan}. Recently, Thomson and Duncan argued that a convective dynamo
mechanism might result in large fields $\sim 10^{15}$G \cite{Tho93}. On the
other hand, it is presumed from the scalar virial theorem \cite{Chan53} based 
on Newtonian gravity that the limiting interior field in neutron stars could be
as large as $\sim 10^{18}$G \cite{Lai}. From the general relativistic 
calculation of axis-symmetric neutron stars in magnetic fields, it follows that
neutron stars could sustain magnetic fields $\sim 10^{18}$G \cite{Boc,Pra00}. 
Because of highly conducting core, such large interior fields may be frozen and
could not be directly accessible to observation. Its effects may be manifested 
in various observables such as the mass-radius
relationship, neutrino emissivity etc. Motivated by the existence of large
fields in the core of neutron stars, its influence on the gross properties of
neutron stars was studied by various groups \cite{Lai,Cha97,Suh,Bro}. 
The calculations in the relativistic mean field (RMF) approach showed that the 
equation of state (EoS) was
modified due to the Landau quantization and also by the interaction of magnetic
moments of baryons with the field \cite{Bro}. The intense magnetic field was 
found to change the composition of beta equilibrated matter relevant to neutron 
stars drastically \cite{Cha97,Suh,Bro}. The neutrino emissivity 
in neutron stars was reported to be enhanced in strong magnetic fields 
\cite{Lei}.

Besides strong interior fields, many exotic forms of matter may exist in the
dense core of neutron stars. One such possibility is the appearance of the 
Bose-Einstein condensate of strange particles. Nelson and Kaplan first pointed
out that antikaons may undergo  the Bose-Einstein condensation (BEC) in dense 
matter at zero temperature because of the attractive s-wave antikaon-nucleon 
interaction
\cite{Kap}. Later, this idea was applied to neutron stars by various authors
\cite{Pra97,Gle99,Pal}. Bose-Einstein condensation in a 
magnetic field is an old and interesting problem in other branches of physics 
also namely condensed matter physics and statistical physics.
It was shown by Schafroth \cite{Sch} that a non-relativistic Bose gas 
could not condense in an external magnetic field. There 
are some calculations on the condensation of relativistic charged Bose
gas in magnetic fields in the literature \cite{Elm,Roj}. Elmfors and 
collaborators \cite{Elm} noted that the relativistic Bose gas might condense 
for spatial dimension $d \geq 5$. They showed that the number density of 
bosons in the ground state diverges for $d < 5$ in the presence of a magnetic
field. On the other hand, it was argued \cite{Roj} that the
condensation of bosons in a magnetic field could occur in three dimension if 
the chemical potential of bosons was taken as a function of density, 
temperature and magnetic field \cite{Roj}. In this case, the BEC 
would be a diffuse one because there is no definite critical temperature. 
It was also shown in the latter calculation \cite{Roj} that 
the number density of bosons in the ground state was finite. Recently,
Suh and Mathews \cite{Suh} have studied pion condensation in a beta 
equilibrated non-interacting n-p-e-$\mu$ system in magnetic fields.

In this paper, we investigate the influence of strong magnetic fields on the
Bose-Einstein condensation of antikaons in cold and dense matter relevant to
neutron stars. This may have profound implications on the gross properties of
neutron stars. This method of studying the BEC in strong magnetic fields and 
dense matter is rather general; therefore it should be of correspondingly 
broad interest.   

We consider strong magnetic field effects on antikaon condensation in the beta
equilibrated neutron star matter composed of neutrons, protons, electrons, 
muons and $K^-$ mesons within the framework of a relativistic field 
theoretical model \cite{Ser}. As the constituents in neutron stars are highly 
degenerate, the chemical potentials of baryons are larger than the temperature
of the system. Therefore, the gross properties of neutron stars are calculated 
at zero temperature. The total Lagrangian density may be written as
the sum of baryonic, kaonic and leptonic parts i.e. ${\cal L} = {\cal L}_B + 
{\cal L}_K + {\cal L}_l$. In a uniform magnetic field, the baryonic Lagrangian
density \cite{Sug} is given by 
\begin{eqnarray}
{\cal L}_B &=& \sum_{B=n,p} \bar\psi_{B}\left(i\gamma_\mu{D^\mu} - m_B
+ g_{\sigma B} \sigma - g_{\omega B} \gamma_\mu \omega^\mu 
- g_{\rho B} 
\gamma_\mu{\mbox{\boldmath $t$}}_B \cdot 
{\mbox{\boldmath $\rho$}}^\mu 
-{\kappa}_B {\sigma_{\mu\nu}} F^{\mu\nu} 
\right)\psi_B\nonumber\\
&& + \frac{1}{2}\left( \partial_\mu \sigma\partial^\mu \sigma
- m_\sigma^2 \sigma^2\right) - U(\sigma) \nonumber\\
&& - \sum_{k=\omega, \rho} \left[ {1\over 4} \left( \partial_{\mu}V_{\nu}^k 
- \partial_{\nu}V_{\mu}^k \right)^2 - {1\over 2} m_k^2 (V_{\mu}^k)^2 \right]
+ \frac{1}{4} g_4 \left( \omega_{\mu} \omega^{\mu}\right)^2  
- {1\over 4} F^{\mu \nu} F_{\mu \nu}~.
\end{eqnarray}
Here $\psi_B$ denotes the Dirac spinor for baryon B with vacuum mass $m_B$
and isospin operator ${\mbox {\boldmath $t$}}_B$. The scalar self-interaction 
term \cite{Bog} is, $U(\sigma) = {g_2 \sigma^3}/3 + {g_3 \sigma^4}/4$.
Following Ref.\cite{Bro}, the interaction of anomalous magnetic moments of 
baryons with magnetic fields is given by the last term under the summation in 
Eq.(1). Here, $F^{\mu\nu}$ is the electromagnetic field tensor, 
$\sigma_{\mu\nu}=[\gamma_{\mu},\gamma_{\nu}]/2$ and ${\kappa}_B$ is the 
experimentally measured value of magnetic moment for baryon $B$.
The covariant derivative for a charged particle is $D^{\mu} = {\partial}^{\mu}
+ i q A^{\mu}$ with the choice of gauge corresponding to the constant magnetic
field ($B_m$) along z-axis is $A_0=0$, ${\bf A} \equiv (0,xB_m,0)$.
The form of 4-component spinor solutions for baryons is given by 
Ref. \cite{Bro}. The (anti)kaon-nucleon interaction is treated in the same
footing as that of the nucleon-nucleon interaction \cite{Gle99}. Therefore, the
kaonic Lagrangian density in a magnetic field is given as,
\begin{equation}
{\cal L}_K = D^*_\mu{K^*} D^\mu K - m_K^{* 2} {K^*} K ~,
\end{equation}
where the covariant derivative 
$D_\mu = \partial_\mu + iq A_{\mu} + ig_{\omega K}{\omega_\mu} + i g_{\rho K} 
{\mbox{\boldmath $t$}}_K \cdot {\mbox{\boldmath $\rho$}}_\mu$. 
There is no interaction term involving magnetic moments in the kaonic 
Lagrangian density because (anti)kaons having zero spin angular momentum do not
possess magnetic moments.
The effective mass of (anti)kaons in this
minimal coupling scheme is given by
$m_K^* = m_K - g_{\sigma K} \sigma$. The solution for negatively
charged kaons in a magnetic field is 
$K \propto \left({qB_{m}/{\pi}}\right)^{1/4} 
({1/{\sqrt{2^{n}n!}}}) {\rm e}^{-i\omega_{K^-} t + ip_y y + ip_z z} 
{\rm e}^{-qB_m{\eta^2}/2} H_n({{\sqrt {qB_m}}\eta})$, 
where $\eta=x + p_y/qB_m$, "H" denotes the Hermite polynomial with $n$ 
the Landau principal quantum number. The Lagrangian density for neutrons is 
obtained by putting $q=0$ in the covariant derivatives of Eq. (1). In the mean 
field approximation \cite{Ser}, the meson field equations in the presence of 
antikaon condensate and magnetic field are 
\begin{eqnarray}
m_\sigma^2\sigma &=& -\frac{\partial U}{\partial\sigma}
+ \sum_B g_{\sigma B} n_B^S 
+ g_{\sigma K} \frac {m^*_K}{\sqrt{m_K^{*2} + q B_m}} n_{K^-}~,\\ 
m_\omega^2\omega_0  + g_4 {\omega_0^3} &=& \sum_B g_{\omega B} n_B
- g_{\omega K} n_{K^-} ~,\\ 
m_\rho^2\rho_{03} &=& \sum_B g_{\rho B} I_{3B} n_B 
+ g_{\rho K} I_{3 K^{-}} n_{K^-} ~.
\end{eqnarray}
where $n_B$ and $n_B^s$ are baryon and scalar density for baryon B 
respectively; $I_{3B}= + 1/2$ for protons, $-1/2$ for neutrons and 
$I_{3 K^-}=-1/2$ for $K^-$ mesons. 
The expressions of the scalar and baryon density corresponding
to protons are given by \cite{Bro} 
\begin{eqnarray}
n^s_p &=& \frac{|q_p| B_m}{2 \pi^2} \sum_{\nu} \sum_s m_p^*
\frac{ \overline{m_p}}{\overline{m_p} - s \kappa_p B_m}
\ln \left( \left| \frac{E_f^p +
k_{f,\nu,s}^p}{ \overline{m_p}} \right| \right) \,,
\end{eqnarray}
and
\begin{equation}
n_p = \frac{|q_p| B_m}{2 \pi^2} \sum_{\nu} \sum_s k_{f,\nu,s}^p \,,
\end{equation}
where the energy spectrum for protons is given by
\begin{eqnarray}
E_{p,\nu,s} &=& \sqrt{ k_z^2 + \left( \sqrt{m_p^{*~2} +
2 \nu q_p B_m}
+ s \kappa_p B_m \right)^2 }
 + g_{\omega_p} \omega_0 + \frac{1}{2} g_{\rho_p} \rho_{03} \,,
\label{ep_kap}
\end{eqnarray}
\begin{equation}
\overline{m_p} = \sqrt{m_p^{*~2} + 2 \nu q_p B_m} 
+ s \kappa_p B_m\,,
\end{equation}
and
\begin{equation}
k_{f,\nu,s}^p = \sqrt{ E_f^{p~2} - \left( \sqrt{m_p^{*~2} 
+ 2 \nu q_p B_m} + s \kappa_p B_m \right)^2} \,.
\end{equation}
Similarly for neutrons, those expressions are given by \cite{Bro}
\begin{equation}
n^s_n = \frac{m_n^*}{4 \pi^2} \sum_s k_{f,s} E_f^n - \overline{m}^2
\ln \left( \left| \frac{ E_f^n + k_{f,s}} {\overline{m}}\right|
\right) \,,
\end{equation}
and 
\begin{equation}
n_n = \frac{1}{2 \pi^2}  \sum_s \frac{1}{3} k_{f,s}^3 +
\frac{1}{2} s \kappa_n B_m \left[ \overline{m} k_{f,s} + E_f^{n~2} \left(
\arcsin \frac{\overline{m}}{E_f^n} - \frac{\pi}{2} \right) \right]\,,
\label{num_kap}
\end{equation}
where
\begin{equation}
E_{n,s} = \sqrt{ k_z^2 + \left( \sqrt{m_n^{*~2} + k_x^2 + k_y^2} +
s \kappa_n B_m \right)^2}  + g_{\omega_n} \omega_0 -
\frac{1}{2} g_{\rho_n} \rho_{03}\,,
\label{en_kap}
\end{equation}
\begin{equation}
\overline{m} = m_n^* + s \kappa_n B_m\,,
\end{equation}
and
\begin{equation}
k_{f,s} = \sqrt{ E_f^{n~2} - \overline{m}^2} \,.
\end{equation}

Solving the equation of motion for antikaons,
the in-medium energy of $K^-$ meson in a magnetic field is obtained as 
$\omega_{K^-} = \sqrt{p_z^2 + m_K^{*2} + qB_m(2n+1)} - g_{\omega K} \omega_0 
- g_{\rho K} \rho_{03}/2$. The condition for the condensation of $K^-$ 
meson in a magnetic field is $p_z=0$ and $n=0$. The number density of $K^-$ 
meson in a magnetic field and in the ground state is obtained from the 
relation $J_{\mu}^K = i(K^*{\partial {\cal L}}/{\partial^{\mu}K^*} 
- {\partial {\cal L}}/{\partial^{\mu}K}~ K)$ and it is given
by, $n_{K^-} = - J^{K^-}_0 = 2 (\omega_{K^-} + g_{\omega K} {\omega}_0 
+ g_{\rho K} \rho_{03}/2) K^* K$.
The total energy density is given by
\begin{eqnarray}
{\varepsilon} &=& \frac{1}{2}m_\sigma^2 \sigma^2 
+ \frac{1}{3} g_2 \sigma^3 
+ \frac{1}{4} g_3 \sigma^4  
+ \frac{1}{2} m_\omega^2 \omega_0^2 + \frac{3}{4} g_4 \omega_0^4 
+ \frac{1}{2} m_\rho^2 \rho_{03}^2  \nonumber \\
&& + \sum_{B=n,p} {\varepsilon}_B 
+ \sum_l{\varepsilon}_l 
+ {\varepsilon}_{\bar K}~,
\end{eqnarray}
where $\varepsilon_B$ and $\varepsilon_l$ correspond to the kinetic energy 
densities of baryons and leptons respectively. The kinetic energy densities 
of protons and neutrons in a magnetic field are given by 
\cite{Bro} 
\begin{equation}
\varepsilon_p = \frac{|q_p| B_m}{4 \pi^2} \sum_{\nu} \sum_s E_f^p k_{f,\nu,s}^p
 + \overline{m_p}^2 \ln \left( \left|
\frac{E_f^p + k_{f,\nu,s}^p}{\overline{m_p}} \right| \right) \,,
\end{equation}
and
\begin{eqnarray}
\varepsilon_n &=& \frac{1}{4 \pi^2} \sum_s \frac{1}{2} E_f^{n~3} k_{f,s}
 + \frac{2}{3} s \kappa_n B_m E_f^{n~3} \left(
\arcsin \frac{\overline{m}}{E_f^n} - \frac{\pi}{2} \right) \nonumber \\
&+& \left( \frac{1}{3} s \kappa_n B_m - \frac{1}{4} \overline{m} \right) \left[
\overline{m} k_{f,s} E_f^n + \overline{m}^3 \ln \left(\left| \frac{ E_f^n 
+ k_{f,s}}
{\overline{m}}\right| \right) \right]\,.
\end{eqnarray}
Similarly, the expression for the kinetic energy density of electrons has the
same form as that of protons but electrons are noninteracting and anomalous
magnetic moment of electrons is not considered here \cite{Bro}.
The energy density for antikaons in the condensate state is 
$\varepsilon_{\bar K} = \sqrt{m^{*2}_K + qB_m} ~ n_{K^-}$. 
The other terms in Eq. (16) represent interaction energy
densities. The pressure of the system follows from the relation 
$P={\mu_n}n_b - \varepsilon$, where 
$\mu_{n}$ and $n_b$ are the neutron chemical 
potential and total baryon density, respectively. In the core of neutron stars, 
strangeness changing processes such as 
$n \rightleftharpoons p + K^-$ and $e^- \rightleftharpoons K^- + \nu_e$ 
occur. The chemical equilibrium yields $\mu_n - \mu_p = \mu_{K^-} = \mu_e$, 
where $\mu_p$ and $\mu_{K^-}$ are respectively the chemical 
potentials of protons and $K^-$ mesons. Employing Eq. (3) in conjunction with 
the chemical equilibrium conditions and charge neutrality 
$n_p - n_{K^-} - n_e - n_{\mu} = 0$, we obtain the effective masses 
self-consistently.

In the effective field theoretical approach adopted here, two different sets of
coupling constants for nucleons and kaons with $\sigma$, $\omega$ and $\rho$
meson are required. The nucleon-meson coupling constants are obtained by 
fitting experimental data for binding energies and charge radii for heavy 
nuclei \cite{Sug}. This set of parameters is known as TM1 set. The values of 
coupling constants are $g_{\sigma N} = 10.0289$, 
$g_{\omega N} = 12.6139$, $g_{\rho N} = 4.6322$, $g_2 = -7.2325 fm^{-1}$,
$g_3 = 0.6183$ and $g_4 = 71.3075$. The incompressibility of matter at normal
nuclear matter density ($n_0=0.145 fm^{-3}$) is 281 MeV for the TM1 model. 
According to the simple quark model and
isospin counting rule, the kaon-vector meson coupling constants are 
$g_{\omega K} = \frac {1}{3} g_{\omega N}$ and $g_{\rho K}=g_{\rho N}$. On the
other hand, the scalar coupling constant is obtained from the real part of the
antikaon optical potential at normal nuclear matter density i.e. 
$U_{\bar K} (n_0) = - g_{\sigma K} {\sigma} - g_{\omega K} {\omega}_0$ 
\cite{Pal}. In this calculation, we have taken $U_{\bar K} (n_0) = -160$ MeV 
and the scalar coupling is $g_{\sigma K} = 2.0098$. 

The TM1 model was adopted earlier for the description of heavy nuclei and the 
equation
of state for neutron stars \cite{Sug}. Besides the non-linear $\sigma$ meson 
terms, the model also includes non-linear $\omega$ meson term. It was shown 
\cite{Sug} that the TM1 model reproduced scalar and vector potentials close to
those of the relativistic Brueckner Hartree Fock calculation using the 
realistic nucleon-nucleon interaction \cite{Brock}. Recently, the TM1 model was
used for the investigation of antikaon condensation in neutron star matter 
\cite{Pal} for zero field. For TM1 parameter set, it was found that the phase 
transition was of second order \cite{Gle99,Pal}.
In the TM1 model, the maximum masses and central densities of neutron stars
without and with antikaon condensation where $U_{K^-}=-160$ MeV are 
respectively, 2.179(1.857)$M_{\odot}$ and 5.97(6.37)$n_0$.

In Figure 1, number densities of various particles are plotted with baryon 
density. The particle densities for $B_m=0$ are shown by the solid lines, 
whereas those corresponding to $B_m = 1.5 \times 10^5 B_c^e$ are denoted by the
dashed lines. The critical field for electrons ($B_c^e$) is that value where
cyclotron quantum is equal to or above the rest energy of an electron and its
value is $B_c^e = 4.414 \times 10^{13}$G. Here we note that the formation of 
$K^-$ condensation is delayed to higher density than the field free case. The 
threshold densities of $K^-$ condensation corresponding to $B_m=0$ and 
$B_m=1.5 \times 10^5 B_c^e$ are 2.67$n_0$ and 3.85$n_0$ respectively. The 
delayed appearance of $K^-$ condensation may be attributed to the stiffer EoS 
because of the effects of magnetic moments. In the presence of the field, 
the enhancement of electron and muon fraction are pronounced whereas the proton 
fraction is smaller than the zero field value beyond 2.7$n_0$.
With the appearance of $K^-$ condensate, it would try to diminish electron and 
muon density. On the other hand, the phase spaces of electrons and muons are so
strongly modified in a quantizing field that their fractions are significantly 
increased. The net result is the reduction in the density of $K^-$ condensate
than that of the field free case. The proton density increases after the onset 
of $K^-$ condensation. The neutron fraction also increases because of the 
interaction of anomalous magnetic moment of neutrons with the field. 
This may have important effects on the equation of state.

In the presence of magnetic fields $> 5 \times 10^{18}$G, the nucleon 
effective mass
is enhanced in the high density regime than that of the field free case. This 
may be attributed to the effects of magnetic moments as it was also noted in 
Ref.\cite{Bro}. The (anti)kaon effective mass in magnetic fields does not
change appreciably from the zero field case. 

The onset of $K^-$ condensation is given by the equality of $K^-$ chemical
potential ($\mu_{K^-}$) with electron chemical potential ($\mu_e$). In the
presence of magnetic field, we find  the hadronic phase smoothly connects 
to the antikaon condensate phase resulting in a second order phase transition 
as it is evident from equation of state (pressure versus energy density curve)
in Figure 2. For TM1 parameter set, we note that the phase transition is of
second order with and without magnetic field.

In Figure 2, matter pressure ($P$) versus matter energy density ($\epsilon$)
is displayed for $B_m=0$ (curve I), $B_m= 4 \times 10^4 B_c^e$ (curve II) and 
$B_m = 1.5 \times 10^5 B_c^e$ (curve III). For $B_m=4 \times 10^4 B_c^e$, 
we note that the curve becomes slightly stiffer with the onset of
$K^-$ condensation. This stiffening may be attributed to the large enhancement
in electron and muon fraction in the field. However, this effect is reduced
in the high density regime where electron and muon fraction become small. As 
the field is further increased to $B_m = 1.5 \times 10^5 B_c^e$, not only 
electrons and muons are strongly Landau quantized, but also protons are 
populated in the zeroth Landau level. It was shown \cite{Cha97,Bro} that Landau
quantization of charged particles was responsible for the softening in the
equation of state. On the other hand, the effects of baryon magnetic moments
for $B_m = 1.5 \times 10^5 B_c^e$ overwhelm the effects of Landau quantization. 
Consequently, the curve corresponding to $B_m = 1.5 \times 10^5 B_c^e$ stiffens 
further. It is found here that the effects of magnetic moments are 
important for $B_m > 10^5 B_c^e$. Besides the effects of Landau quantization
and magnetic moments, the contribution of electromagnetic field to the matter
energy density and pressure is to be taken into account. The magnetic energy 
density and pressure, ${\varepsilon_f}=P_f=B_m^2/(8\pi) = 4.814 \times 10^{-8} 
(B_m/B_c^e)^2 MeV fm^{-3}$, become significant in the core of the star for 
$B_m \geq 10^5 B_c^e$. 

In this calculation, we have considered interior magnetic field 
$> 5 \times 10^{18}$G. However, it was found in a recent calculation \cite{Bro2}
that the maximum value of the magnetic field within a star may not exceed 
$3 \times 10^{18}$G for a particular choice of a constant current function
but independent of an EoS. In this case, the ratio of the maximum field to the
average field is not large because of small spatial gradient. The authors
\cite{Bro2} argued that the value of maximum field at any point may well 
exceed the average value as mentioned above for a different field geometry. In
that event the effects of strong magnetic field $> 5 \times 10^{18}$G on the
threshold of antikaon condensation, particle composition and EoS might be
important.  
 
To summarise, in this paper, we have focused on the formation of the antikaon 
condensation 
in dense nuclear matter in the presence of  magnetic fields. We have considered
the interaction of magnetic moments of baryons with the field and the magnetic 
energy density and pressure in this work. In the presence of strong magnetic 
fields $> 5 \times 10^{18}$G, we find a considerable change in the phase 
spaces of charged particles. The threshold density of $K^-$ condensation 
is delayed to higher density in the presence of such a strong field and the EoS 
becomes stiffer.  For $B_m > 10^{18}$G, the effects of magnetic 
moments are 
important and it adds to further stiffening of the equation of state. Also, the 
electromagnetic field contribution to the energy density and pressure becomes
important in the core for $B_m > 10^{18}$G. 
The stiffening of the EoS in the presence of 
magnetic fields might have significant impact on the gross properties of 
neutron stars such as the mass-radius relationship, cooling etc. It is worth 
mentioning here that (anti)kaons do not interact with magnetic fields in the 
same way as fermions do because their spin angular momentum is zero. 

In this calculation, we do not include the role of hyperons, pion condensation 
and 
nucleon-nucleon correlation on the antikaon condensation. Negatively charged
hyperons, in particular $\Sigma^-$ hyperon, could delay the onset of $K^-$
condensation \cite{Pra97}. However, it was estimated that $\Sigma^-$-nucleon
interaction is highly repulsive in normal nuclear matter \cite{Fri}. Recently,
it has been also shown that threshold densities of most hyperons including 
$\Sigma^-$ are
substantially increased in strong magnetic field $B_m > 5 \times 10^{18}$G both
due to Landau quantisation and magnetic moment interactions \cite{Bro2}. In this
situation, $\Sigma^-$ hyperons might have no
impact on $K^-$ condensation. Pion condensation could occur in neutron stars 
because of the attractive $p$-wave pion-nucleon interaction \cite{Akm}. 
The condensation of $\pi^-$ may modify the electron chemical potential which, 
in turn, would delay $K^-$ condensation. In this paper, we have employed 
the RMF model which does not include nucleon-nucleon correlations. It was shown
in non relativistic models \cite{Pan} that nucleon-nucleon correlations shifted
the threshold density of $K^-$ condensation to higher density. We believe that 
the qualitative features of strong magnetic fields presented 
here would survive even in other models which include 
hyperons, pion condensation and nucleon-nucleon correlations. It would be 
interesting to look into the neutrino emissivity from an antikaon condensed 
matter and the structure of compact stars having antikaon condensate in the 
presence of a strong magnetic field. 


\newpage
\vspace{-6cm}

\epsfxsize=15cm
\epsfysize=20cm
\epsffile{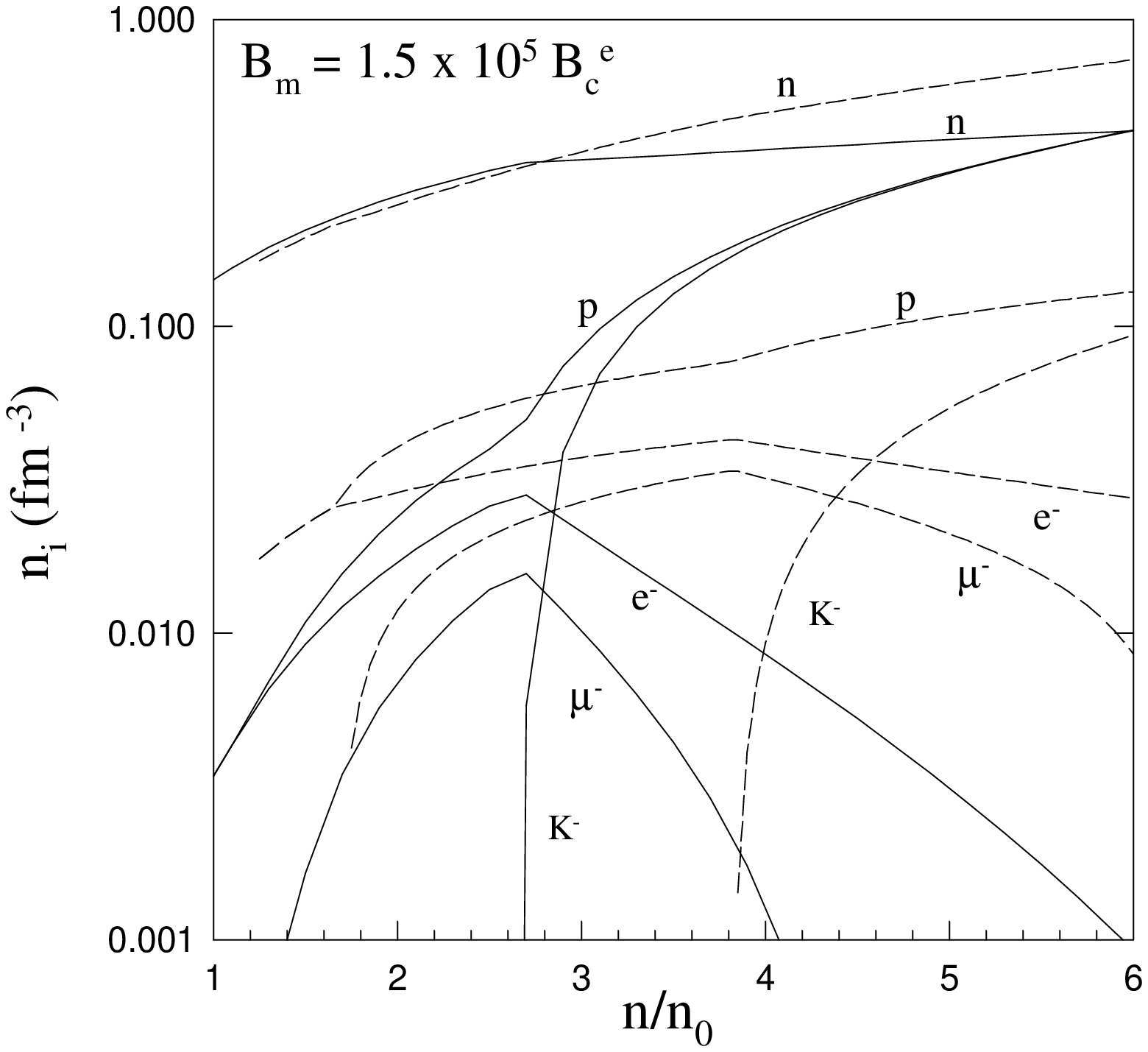}
\noindent{Fig.1 : The particle abundances are plotted with normalised 
baryon density
for $B_m=0$ and $B_m = 1.5 \times 10^5 B_c^e$. Solid lines indicate particle
abundances for field free case whereas dashed lines denote those with the
magnetic field. The critical electron field ($B_c^e$) is 
$4.414 \times 10^{13}$G.} 


\newpage
\vspace{-6cm}

\epsfxsize=15cm
\epsfysize=20cm
\epsffile{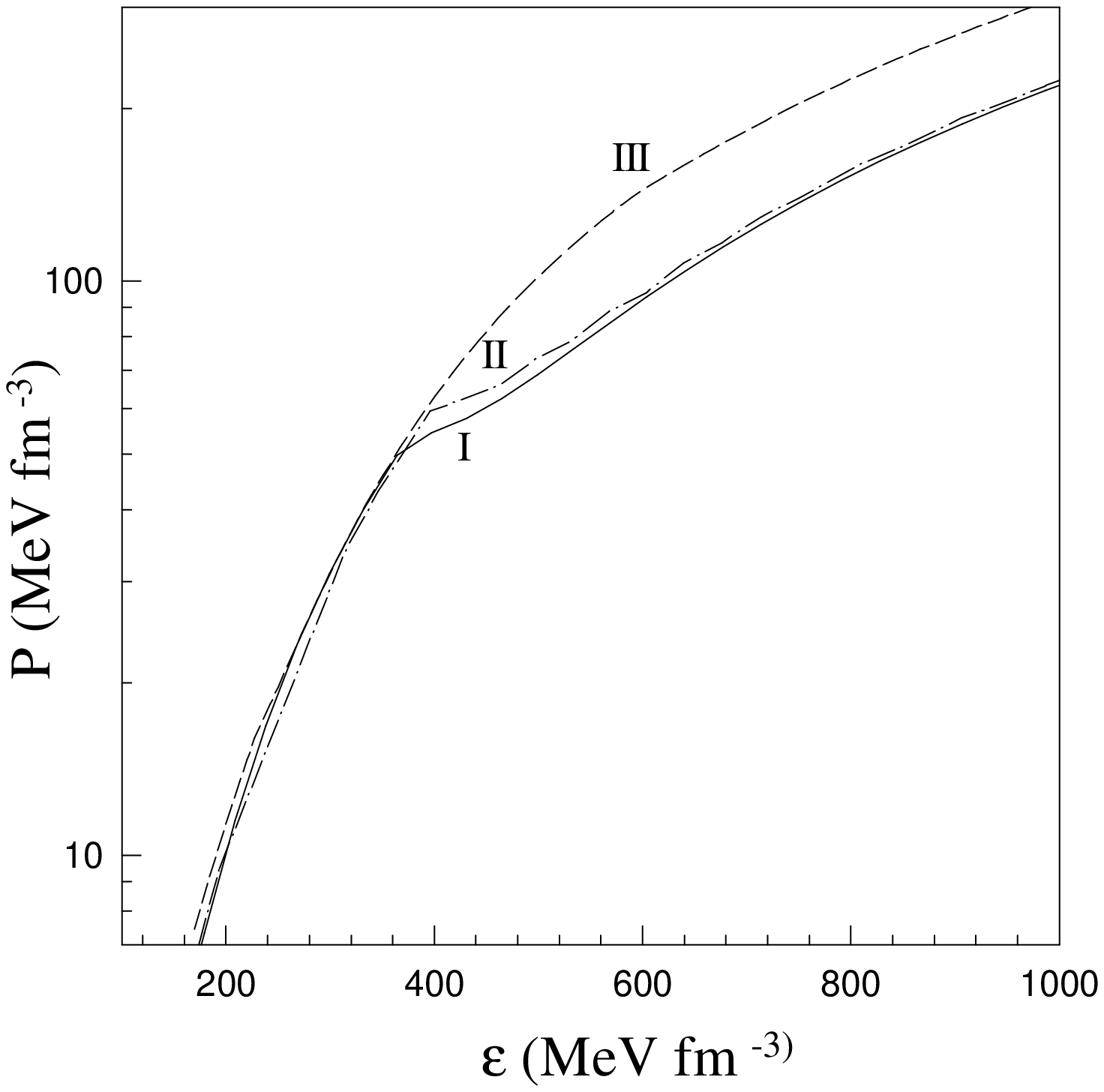}
\noindent{Fig.2 : The matter pressure ($P$) is shown as a function of 
matter energy density ($\varepsilon$) for different values of $B_m$. The field
free case is shown by curve I (solid line) and curve II (dash-dotted line) and
curve III (dashed line) represent calculations for $B_m = 4 \times 10^4 B_c^e$
and $B_m = 1.5 \times 10^5 B_c^e$, respectively. The critical electron field
($B_c^e$) is $4.414 \times 10^{13}$G.}  


\newpage

\begin{thebibliography}{99}

 
\bibitem{Kou98} Kouveliotou C, Dieter S, Strohmayer T, Paradijs J van,
Fishman G J, Meegan C A and Hurley K 1998 Nature {\bf 393} 235; 
Kouveliotou C, Strohmayer T, Hurley K, Paradijs J van,
Finger M H, Dieter S, Woods P, Thompson C and Duncan R C 1999 Astrophys. J.
{\bf 510} L115; Heyl J S and Kulkarni S R, 1998 Astrophys. 
J. {\bf 506} L61; Chatterjee P, Hernquist L and Narayan R, 
2000 Astrophys. J. {\bf 534} 373.
\bibitem{Dun92} Duncan R C and Thompson C 1992 Astrophys. J. {\bf 392} L9;
Thompson C and Duncan R C 1995 MNRAS {\bf 275} 255;
Thompson C and Duncan R C 1996 Astrophys. J. {\bf 473} 322;
Usov V V 1992 Nature {\bf 357} 472;
Paczynski B 1992 Acta. Astron. {\bf 42} 145;
Vasisht G and Gotthelf E V 1997 Astrophys. J. {\bf 486} L129.
\bibitem{Chan} Chanmugam G 1992 Annu. Rev. Astron. Astrophys. {\bf 30} 143.
\bibitem{Tho93} Thompson C and Duncan R C 1993 Astrophys. J. {\bf 498} 194.
\bibitem{Chan53} Chandrasekhar S and Fermi E 1953 Astrophys. J. {\bf 118} 116.
\bibitem{Lai} Lai D and Shapiro S L 1991 Astrophys. J. {\bf 383} 745.
\bibitem{Boc} Bocquet M, Bonazzola S, Gourgoulhon E and Novak J 1995 Astron.
Astrophys.{\bf 301} 301.
\bibitem{Pra00} Cardall C Y, Prakash M and Lattimer J M 2001 Astrophys. J.
{\bf 554} 322.
\bibitem{Cha97} Chakrabarty S, Bandyopadhyay D and Pal S 1997 Phys. Rev. Lett. 
{\bf 78} 2898; Bandyopadhyay D, Chakrabarty S and Pal S 1997 Phys. Rev. 
Lett. {\bf 79} 2176.
\bibitem{Suh} Suh I -S and Mathews G J 2001 Astrophys. J. {\bf 546} 1126.
\bibitem{Bro} Broderick A, Prakash M and Lattimer J M 2000 Astrophys. J.
{\bf 537} 351.
\bibitem{Lei} Leinson L B and Perez A 1998 JHEP {\bf 09} 020;
Bandyopadhyay D, Chakrabarty S, Dey P and Pal S 1998 Phys. Rev. D{\bf 58} 
121301;
Baiko D A and Yakovlev D G 1999 Astron. Astrophys. {\bf 342} 192; 
Dalen E N E van, Dieperink A E L, Sedrakian A and 
Timmermans R G E 2000 Astron. Astrophys. {\bf 360} 549. 
\bibitem{Kap} Kaplan D B and Nelson A E 1986 Phys. Lett. B{\bf 175} 57;
Nelson A E and Kaplan D B 1987 Phys. Lett. B{\bf 192} 193.
\bibitem{Pra97} Prakash M, Bombaci I, Prakash M, Ellis P J, 
Lattimer  J M and Knorren R 1997 Phys. Rep. {\bf 280} 1.
\bibitem{Gle99} Glendenning N K and Schaffner-Bielich J 
1998 Phys. Rev. Lett. {\bf 81} 4564;
Glendenning N K and Schaffner-Bielich J 1999 Phys. Rev.
C{\bf 60} 025803.
\bibitem{Pal} Pal S, Bandyopadhyay D and Greiner W 2000 Nucl. Phys. A{\bf 674}
553; Banik S and Bandyopadhyay D 2001 Phys. Rev. C{\bf 63} 035802.
\bibitem{Sch} Schafroth M R 1955 Phys. Rev. {\bf 100} 463.
\bibitem{Elm} Elmfors P et al. 1995 Phys. Lett. B{\bf 348} 462 and 
references therein.
\bibitem{Roj} Perez R H 1996 Phys. Lett. B{\bf 379} 148 and references
therein.
\bibitem{Ser} Serot B D and Walecka J D 1986 Adv. Nucl. Phys. {\bf 16} 1.
\bibitem{Sug} Sugahara Y and Toki H 1994 Nucl. Phys. A{\bf 579} 557.
\bibitem{Bog} Boguta J and Bodmer A R 1977 Nucl. Phys. A{\bf 292} 413.
\bibitem{Brock} Brockmann R and Machleidt R 1990 Phys. Rev. C{\bf 42} 1965.
\bibitem{Bro2} Broderick A, Prakash M and Lattimer J M 2001 astro-ph/0111516
\bibitem{Fri} Friedman E, Gal A, Mares J and Cieply A 1999 Phys. Rev. C{\bf 60}
024314.
\bibitem{Akm} Akmal A, Pandharipande V R and Ravenhall D G 1998 
Phys. Rev. C{\bf 58} 1804.
\bibitem{Pan} Pandharipande V R, Pethick C J, Thorsson V 1995 Phys. Rev. Lett. 
{\bf 75} 4567; Carlson J, Heiselberg H and Pandharipande V R 2001 Phys. Rev. 
C{\bf 63} 017603. 

\end{thebibliography}
\end{document}